\documentclass[aps,twocolumn,prb,groupedaddress,amsmath,amssymb]{revtex4-2}
\usepackage{dcolumn}
\usepackage{bm}
\usepackage{amsmath}
\usepackage{txfonts}
\usepackage[T1]{fontenc}
\usepackage{xspace}
\usepackage{ulem}
\usepackage{comment}
\usepackage{braket}
\ifx\pdfoutput\undefined
\usepackage[dvipdfmx]{graphicx}
\usepackage[dvipdfmx]{hyperref}
\usepackage[dvipdfmx]{color}
\else
\usepackage{graphicx}
\usepackage{hyperref}
\usepackage{color}
\fi

\graphicspath{
{./image/}
}
\begin{document}

\title{Elliptical-rod geometries enhance photonic band gaps in disordered stealthy hyperuniform photonic crystals}
\author{Kota Asakura}
\email{asakura@stat.phys.titech.ac.jp}
\author{Kazuki Yamamoto}
\author{Akihisa Koga}
\affiliation{Department of Physics, Institute of Science Tokyo,
  Meguro, Tokyo 152-8551, Japan}

\date{\today}

\begin{abstract}
  We study two-dimensional photonic crystals
  composed of elliptical dielectric rods
  arranged according to
  stealthy hyperuniform point patterns.
  These patterns are characterized by the structure factor,
  which vanishes for $0 < |\bm{k}| \le K$,
  where $\bm{k}$ is the wave number and
  $K$ denotes the cutoff wave number specifying
  the stealthiness of the pattern.
  The optical properties of the photonic crystals are analyzed
  by applying the plane-wave expansion method to Maxwell's equations.
  We demonstrate that photonic crystals composed of elliptical dielectric rods
  can exhibit larger photonic band gaps
  than those with cylindrical rods when both the rod orientation
  and aspect ratio are properly optimized.
  This behavior contrasts with that of periodic lattices
  such as triangular or square arrays.
  These findings shed light on the crucial role of structural anisotropy and
  aperiodic structure in enhancing photonic band-gap formation.
\end{abstract}

\maketitle
\section{Introduction}\label{sec_intro}

A photonic crystal, composed of alternating materials with different dielectric constants~\cite{Ho90},
exhibits a photonic band gap (PBG) that prohibits light propagation within certain frequency ranges~\cite{John87,Yablonovitch87},
enabling applications~\cite{Altug06} such as lasers~\cite{Noda01,Notomi03}, sensors~\cite{El-Kady06},
and integrated optical circuits~\cite{Chutinan03}.
More recently, the study of photonic band structures has provided
a powerful framework for investigating cutting-edge topics
like topological photonics~\cite{Lu14,Ozawa19} and
non-Hermitian physics~\cite{Feng17}.
So far, photonic crystals have been mainly investigated
in periodic~\cite{Chan91,Datta92,Dyogtyev10,Lawrence09} and
quasiperiodic~\cite{Zoorob00,Rechtsman08,Florescu09Oct,Notomi04,Takemori_2025}
structures, but these architectures suffer from low tolerance to fabrication imperfections and exhibit a strong directional dependence in their light propagation properties.
Therefore, there is a growing demand for structures that
can exhibit isotropic characteristics and achieve more robust PBGs.

Disordered hyperuniform point patterns~\cite{Torquato03,Torquato18} 
-- characterized by suppressed long-wavelength density fluctuations and a structure factor 
satisfying $\lim_{\bm k \to 0}S(\bm k)=0$ --
have attracted considerable attention 
as candidate structures for next-generation photonic materials 
with their isotropic and robust PBGs.
Such patterns have been observed
in a variety of fields such as
soft matter~\cite{Berthier11,Kurita11}, solids~\cite{Zheng16,Llorens20},
active matter~\cite{Huang21}, biology~\cite{Jiao14},
and cosmology~\cite{Gabrielli02}.
A particularly useful subclass is the stealthy hyperuniform system, defined by
$S(\bm k)=0$ for $0<|\bm k|\le K$~\cite{Batten08,Leseur16},
where $K$ is a certain cutoff parameter.
The degree of stealthiness is quantified by the parameter $\chi$, 
representing the fraction of constrained wavevectors. 
It has been clarified that increasing $\chi$
reduces fluctuations in point spacing and
leads to a more uniform point distribution,
yielding photonic crystals with a wider PBG~\cite{Florescu09Jan}.


Importantly, novel photonic crystal structures such as tri-ellipse configurations \cite{Wen08} and hybrid patterns combining cylindrical and elliptical holes \cite{Janrao12,Sharma16} have been investigated. These studies demonstrate that, by reducing lattice symmetry and introducing geometric anisotropy, it is possible to achieve large complete PBGs supporting both transverse electric (TE) and transverse magnetic (TM) modes, as well as improved slow-light performance and tailored dispersion properties. Moreover, it has been reported that replacing statistically isotropic disordered hyperuniform point patterns with cylindrical dielectric rods can enhance the TM mode of PBGs \cite{Wiersma13,Florescu09Jan}.
Although cylindrical dielectric rods are used in many photonic crystals, 
it remains unclear whether this specific geometry is the most effective for the formation of PBGs. 
While PBGs in stealthy hyperuniform structures have been extensively explored \cite{Florescu09Jan,FroufePere16}, the impact of elliptical dielectric rods on it has received little attention.

In this study, we investigate two-dimensional photonic crystals composed of
elliptical dielectric rods arranged
according to stealthy hyperuniform point patterns.
Calculating dispersion relations for the TM mode
in terms of the plane-wave expansion method,
we demonstrate that, when rod orientation and aspect ratios
are appropriately optimized,
larger PBGs appear compared to the case of cylindrical rods.
This work provides a design strategy for photonic crystals
with enhanced PBGs, 
thereby broadening potential applications in photonic waveguides and
photonic integrated circuits.

The rest of this paper is organized as follows.
In Sec.~\ref{sec_methods}, we explain the procedure for generating
stealthy hyperuniform point patterns.
Then, we describe the plane-wave expansion method for solving
Maxwell's equations in photonic crystals.
In Sec.~\ref{sec_results},
we examine optical properties of photonic crystals
where elliptical dielectric rods are arranged
according to stealthy hyperuniform point patterns.
Finally, conclusions are given in Sec.~\ref{sec_conclusion}.

\section{Model and method}\label{sec_methods}
In this section, we give a method for generating hyperuniform point patterns.
We explain the plane-wave expansion method
to examine optical properties of
the photonic crystals~\cite{Joannopoulos95,Johnson01}.

\subsection{Stealthy hyperuniform point patterns}

Here, we consider the spatial structure of the points
in the rectangular region of size $L_x\times L_y$~\cite{Uche06}.
Its structure factor is defined as
\begin{align}
  S (\bm{k}) &=\frac{1}{N}|\rho(\bm{k})|^2,\label{sofk}\\
  \rho(\bm{k})&= \sum_i e^{i \bm{k}\cdot \bm{r}_i} \label{ck}
\end{align}
where $N$ is the number of points in the system, $\bm{k}$ is the wave number,
and $\rho(\bm{k})$ is the Fourier transform of the point distribution.
%
When periodic boundary conditions are imposed on the unit cell,
the allowed component of $\bm{k}$ is given by
\begin{equation}\label{kg}
  k_\gamma = \frac{2 \pi n_\gamma}{L_\gamma},
\end{equation}
where $\gamma=x,y$ and $n_\gamma$ is an integer.

To generate a point pattern with a certain structure factor,
we introduce the objective function $\Phi\:(\ge 0)$, given by
\begin{equation}
  \Phi(\bm r_1,\bm r_2,\dots,\bm r_N) = \sum_{\bm k} V(\bm k)\Big[S(\bm k) - S_{\mathrm{target}}(\bm k)\Big]^2,
  \label{eq_phi}
\end{equation}
where $S_{\mathrm{target}}(k)$ is the target structure factor and
$V(k)\:[> 0]$ is the window function.
When the objective function is minimized with
adjusting the point patterns $\{\bm{r}_i\}$,
one can obtain the point pattern with the target structure factor.

In this study, we focus on the stealthy hyperuniform point patterns,
which are characterized by
$S(\bm{k})=0$ for $0<|\bm{k}|\le K$.
The target structure factor and window function are given as
\begin{align}
  S_{\mathrm{target}}(\bm k) &= 0,\\
  V(\bm k) &=
  \begin{cases}
    1 & (|\bm k| \leq K) \\
    0       & (|\bm k| > K)
  \end{cases},
\end{align}
where $K$ is some positive number.
Now, we introduce the stealthiness parameter $\chi$
to represent the ratio of the number of constrained degrees of freedom
relative to the total number of degrees of freedom, as
\begin{equation}\label{ch}
  \chi = \frac{M_K}{dN},
\end{equation}
where $M_K\:[=(N_K-1)/2]$, $d$ indicates the spatial dimension,
and $N_K$ is the number of $k$ points with $V(\bm{k})=1$.

\begin{figure}[htb]
  \includegraphics[width=\linewidth]{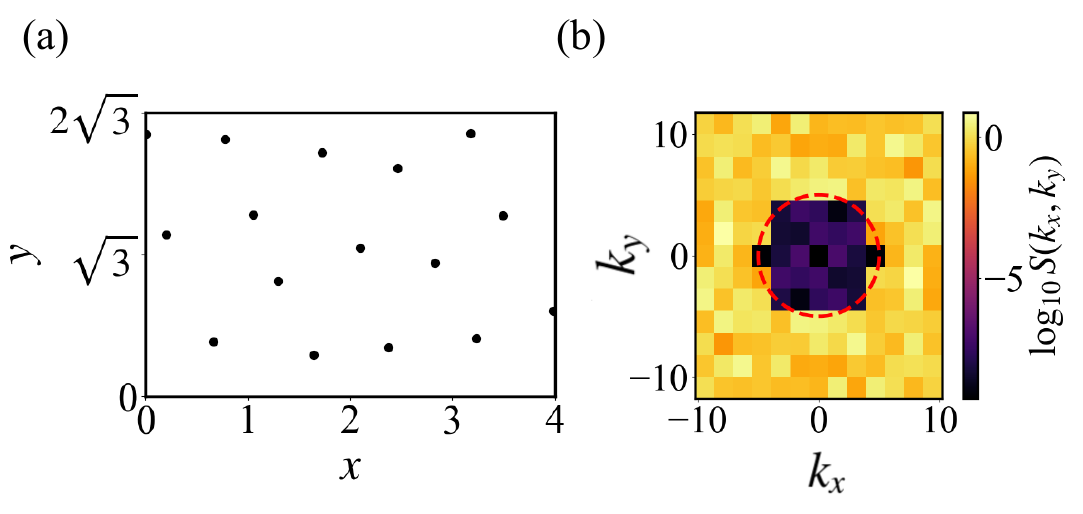}
  \caption{
    (a) Stealthy hyperuniform point pattern with $\chi=0.41$ in the system with $N=16$ and $(L_x, L_y)=(4, 2\sqrt{3})$
    and (b) its structure factor, plotted on a logarithmic scale. Dashed line indicates the boundary of the window function.
  }
  \label{fig1}
\end{figure}
By using the above scheme with $K=5.0$ and $N=16$
in the system with $L_x=4$ and $L_y=2\sqrt{3}$,
we obtain a stealthy hyperuniform point pattern.
Figure~\ref{fig1} illustrates the resulting point configuration with $\chi=0.41$ and its corresponding structure factor \( S(\bm{k}) \).
We find that $S(k)$ is almost zero within the circular region $0\le \bm{k} \le K$.
%
Previous studies have reported that 
stealthy hyperuniform point patterns with larger $\chi$
are promising for applications in photonic crystals~\cite{Florescu09Jan}.

\subsection{Optical properties}
We briefly review the plane-wave expansion method
to analyze optical properties of photonic crystals.
The Maxwell's equations for electromagnetic waves are given by
\begin{equation} \label{1}
  \nabla \times \bm{E} = -\mu_0 \mu \frac{\partial \bm{H}}{\partial t},
\end{equation}
\begin{equation} \label{2}
  \nabla \times \bm{H} = \varepsilon_0 \varepsilon(\bm r) \frac{\partial \bm{E}}{\partial t},
\end{equation}
\begin{equation}
  \nabla \cdot \bm{E} = 0,
\end{equation}
\begin{equation}
  \mu_0 \mu \nabla \cdot \bm{H} = 0,
\end{equation}
where $\bm{E}$ ($\bm{H}$) is the electric (magnetic) field,
$\mu_0$ ($\varepsilon_0$) is the vacuum permeability (permittivity),
and 
$\mu$ [$\varepsilon(\bm r)$] is
the relative permeability (permittivity).
We assume that, in the photonic crystal,
the relative permittivity $\varepsilon (\bm r)$ is position-dependent,
and the permeability is spatially independent.
When the magnetic field oscillates with frequency $\omega$,
$\bm{H}(\bm r, t) = \bm{H}(\bm r) \exp(-i \omega t)$,
we obtain
\begin{equation} \label{ht}
  \nabla \times \left( \frac{1}{\varepsilon(\bm r)} \nabla \times \bm{H}(\bm r) \right) = \frac{\omega^2}{c^2} \bm{H}(\bm r),
\end{equation}
where $c$ is the speed of light in vacuum.
Since the electric field is determined by the magnetic field, as
\begin{equation}
  \bm{E}(\bm{r})=-\frac{1}{i\omega\varepsilon_0\varepsilon(\bm{r})}\nabla \times \bm{H}(\bm{r}),
\end{equation}
we focus on the magnetic field in the following.
In this study, the periodic boundary condition is imposed on the system.
$1/\varepsilon(\bm r)$ can be expanded
using the reciprocal lattice vectors \(\bm{G}\) as
\begin{equation} \label{ep}
  \frac{1}{\varepsilon(\bm r)} = \sum_{\bm{G}} \tilde{\varepsilon}(\bm{G}) \exp(i \bm{G} \cdot \bm r).
\end{equation}

Now, we make use of the plane-wave expansion to the magnetic field.
When one considers the plane wave with a wave number $\bm{k}$,
the magnetic field is rewritten as,
\begin{align}
\label{H}
  \bm{H}(\bm r)
  &= \sum_{\bm{G}} \sum_{\lambda=1}^{2} h_{\bm{G}}^{\lambda} \bm{e}_{\bm{G}}^{\lambda} \exp(i (\bm k + \bm{G}) \cdot \bm r),
\end{align}
where 
$h_{\bm{G}}^{\lambda}$ ($\bm{e}_{\bm{G}}^{\lambda}$) is the amplitude (unit vector)
for the $\lambda$th mode,
and $\lambda$ represents degrees of freedom for the polarization.
We note that $H(r)$, $h_G^\lambda$, and $e_G^\lambda$
depend on the wave number $k$.
By substituting Eqs.~\eqref{ep} and \eqref{H} into Eq.~\eqref{ht},
we obtain the eigenvalue equation as
\begin{align}
\label{ei}
  \sum_{\bm{G}'} & \tilde{\varepsilon}(\bm{G} - \bm{G}') |\bm k + \bm{G}| |\bm k + \bm{G}'|\;
\hat{M}_{GG'}\;
  \begin{pmatrix}
    h_{\bm{G}'}^{1} \\[3pt]
    h_{\bm{G}'}^{2}
  \end{pmatrix} 
  = \frac{\omega^2}{c^2}
  \begin{pmatrix}
    h_{\bm{G}}^{1} \\[3pt]
    h_{\bm{G}}^{2}
  \end{pmatrix},
\end{align}
with
\begin{align}
  \hat{M}_{GG'}=
  \begin{pmatrix}
    \bm{e}_{\bm{G}}^{2} \cdot \bm{e}_{\bm{G}'}^{2} & -\bm{e}_{\bm{G}}^{2} \cdot \bm{e}_{\bm{G}'}^{1} \\[3pt]
    -\bm{e}_{\bm{G}}^{1} \cdot \bm{e}_{\bm{G}'}^{2} & \bm{e}_{\bm{G}}^{1} \cdot \bm{e}_{\bm{G}'}^{1}
  \end{pmatrix}.
\end{align}
In two-dimensional systems,
the polarization directions can be chosen as
$\bm{e}_{\bm{G}}^{1} = (0, 0, 1)$ and 
$\bm{e}_{\bm{G}}^{2} = \frac{1}{|\bm k + \bm{G}|}(k_y + G_y, -(k_x + G_x), 0)$,
without loss of generality.
Since $\bm{e}_{\bm{G}}^{1}\cdot \bm{e}_{\bm{G}}^{2}=0$,
the eigenvalue equation \eqref{ei} decouples into two independent parts.
Namely,
\(h_{\bm{G}}^{1}\) represents the magnetic field in the \(z\)-direction
while the electric field oscillates in the \(xy\)-plane.
Therefore, this mode is referred to as the TE mode.  
On the other hand, \(h_{\bm{G}}^{2}\) represents the magnetic field
in the \(xy\)-plane and the corresponding mode is called the TM mode. 
Thus, the TE and TM modes can be treated independently.

The equations for the TE and TM modes are given as
\begin{align}
  &\sum_{\bm{G}'} \tilde{\varepsilon}(\bm{G} - \bm{G}') \, (\bm k + \bm{G}) \cdot (\bm k + \bm{G}') \, h_{\bm{G}'}^{1}
  = \frac{\omega^2}{c^2} h_{\bm{G}}^{1},\\
  &\sum_{\bm{G}'} \tilde{\varepsilon}(\bm{G} - \bm{G}') \, |\bm k + \bm{G})| |\bm k + \bm{G}'| \, h_{\bm{G}'}^{2}
  = \frac{\omega^2}{c^2} h_{\bm{G}}^{2}.\label{eigm}
\end{align}
In principle, these equations lead to an eigenvalue problem
which contains an infinite set of reciprocal-lattice vectors $\bm{G}$.
In practice, however, the low-energy band structures,
we focus on in the study, can be evaluated
by considering only a finite number of $G$ vectors
in the vicinity of the origin.
The number of $G$ vectors taken into account determines
the matrix dimensions, and
diagonalizing this matrix provides the corresponding dispersion relations.

In the following analysis,
we restrict our discussion to the TM mode
since the photonic band structures for TE and TM modes differ significantly.
Moreover, we focus on photonic crystals composed of dielectric rods in air,
where large band gaps are expected
in the dispersion relations for the TM mode~\cite{Matthews05,Florescu09Jan}.
To discuss optical properties of realistic systems,
we examine a two-dimensional photonic crystal composed of silicon rods in air,
where the dielectric constants are taken as
$\varepsilon = 11.9$ for the silicon and 
$\varepsilon = 1$ for the air.
In our study, the number of rods per unit cell
is fixed at 16 for a systematic analysis.

\begin{figure}[htb]
  \centering
  \includegraphics[width=\linewidth]{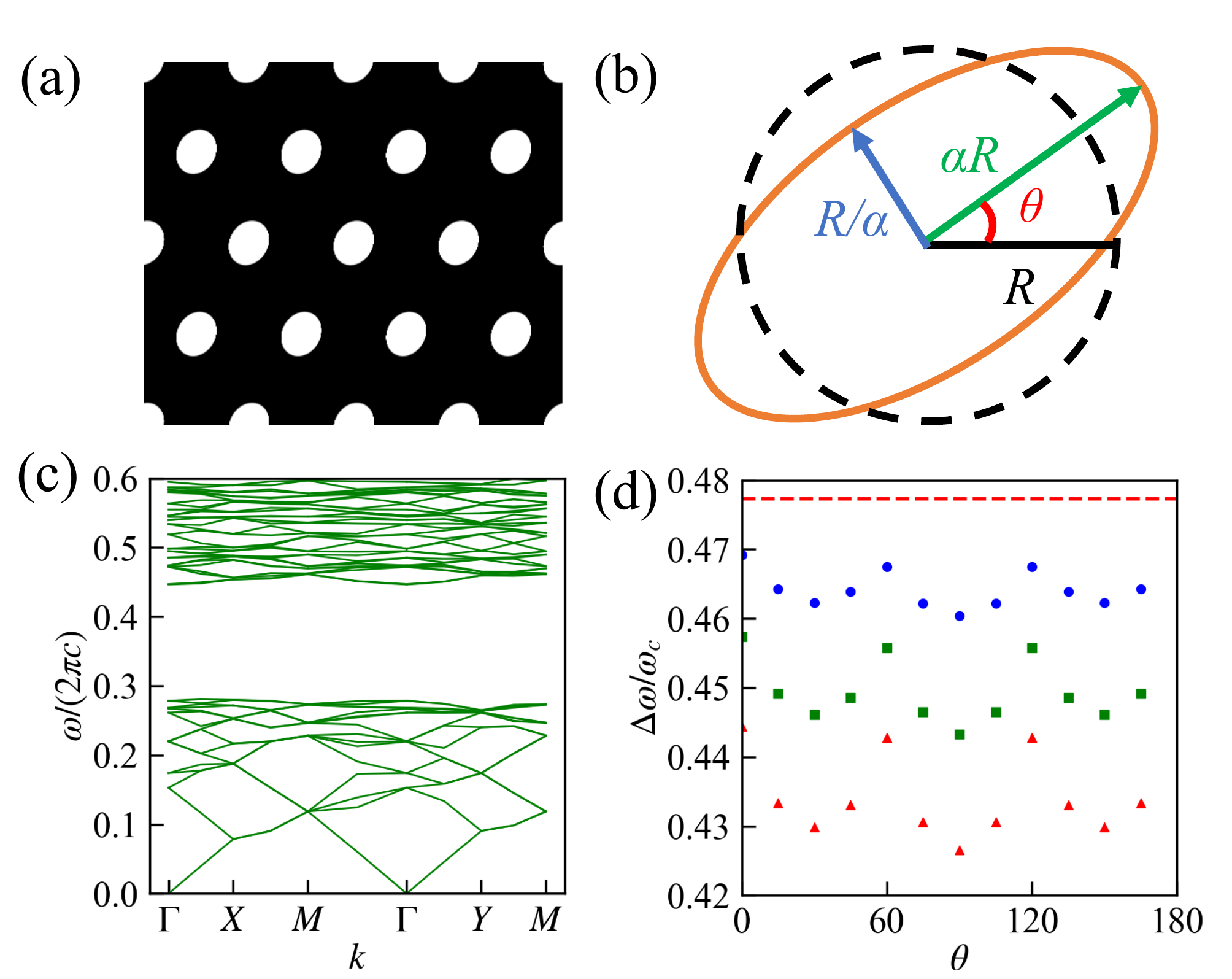}
  \caption{
    (a) Dielectric profile for the photonic crystal composed of the elliptrical rods
    on the triangular lattice with $\alpha=1.1$ and $\theta=60^{\circ}$.
    (b) Cylindrical and elliptical rods.
    (c) Dispersion relation for the photonic crystal with $\alpha=1.1$ and $\theta=60^{\circ}$.
    (d) Relative PBG as a function of $\theta$ in the system with $\alpha = 1.0$ (red dashed line), $1.05$ (blue circles), $1.1$ (green squares), and $1.15$ (red triangles).
}
  \label{fig2}
\end{figure}
\newpage
Before discussing photonic crystals with stealthy hyperuniform structures,
we first examine optical properties of photonic crystals
composed of periodically arranged elliptical rods.
As a simple example, we consider a photonic crystal based on a triangular lattice,
as shown in Fig.~\ref{fig2}(a).
We define a rectangle region with $L_x=4$ and $L_y=2\sqrt{3}$,
containing 16 lattice points.
In this case, the lattice constant is unity.
Each elliptical rod placed on a vertex site has 
three degrees of freedom -- the lengths of the major and minor axes and the orientation angle --
as shown in Fig.~\ref{fig2}(b).
We introduce a deformation parameter $\alpha_i\:(\ge1)$
to describe the $i$th ellipse having the same area as a circle of radius $R$,
such that
the semi-major and semi-minor axes are given by
$\alpha_i R$ and $R/\alpha_i$, respectively.
Here, we consider the photonic crystal where all elliptical rods
share the same values of $\alpha$ and $\theta$, with $R=0.2$
[see Fig.~\ref{fig2}(a)].

The dispersion relation for the TM mode in the photonic crystal composed of the elliptical rods
with $\alpha=1.1$ and $\theta=60^{\circ}$ is shown in Fig.~\ref{fig2}(c). 
We find that the dispersion relation is absent in the energy range
between $\omega_l \sim 0.28\times 2\pi c$ and
$\omega_u \sim 0.45\times 2\pi c$,
where $\omega_l \:(\omega_u)$ is the lower (upper) edge of
the photonic band gap.
Therefore, the PBG is obtained as $\Delta\omega=\omega_u-\omega_l\:
(\sim 0.17\times 2\pi c)$.
To clarify the role of the elliptical rods for the PBG formation,
we show in Fig.~\ref{fig2}(d) the normalized PBG $\Delta\omega/\omega_c$
as a function of $\theta$ for the system with $\alpha=1.0, 1.05, 1.1$, and $1.15$,
where $\omega_c\:[=(\omega_u+\omega_l)/2]$ denote the center of the gap.
In the following, we refer to $\Delta \omega/\omega_c$ as the relative PBG.
We find that the relative PBG of the photonic crystal with elliptical rods $(\alpha\neq 1)$ does not exceed
that with cylindrical rods $(\alpha=1)$ for any value of $\alpha$ or $\theta$ on the triangular lattice.
This result indicates that the introduction of ellipticity does not enhance the PBG
in the triangular-lattice system.
A Similar tendency is observed for the square-lattice system (not shown).
However, it remains nontrivial whether such behavior also persists
in the systems with stealthy hyperuniform point patterns
since those are structurally disordered.

\section{Photonic band gap in stealthy hyperuniform structures}
\label{sec_results}

\begin{figure}[hbt]
  \centering
\includegraphics[width=\linewidth]{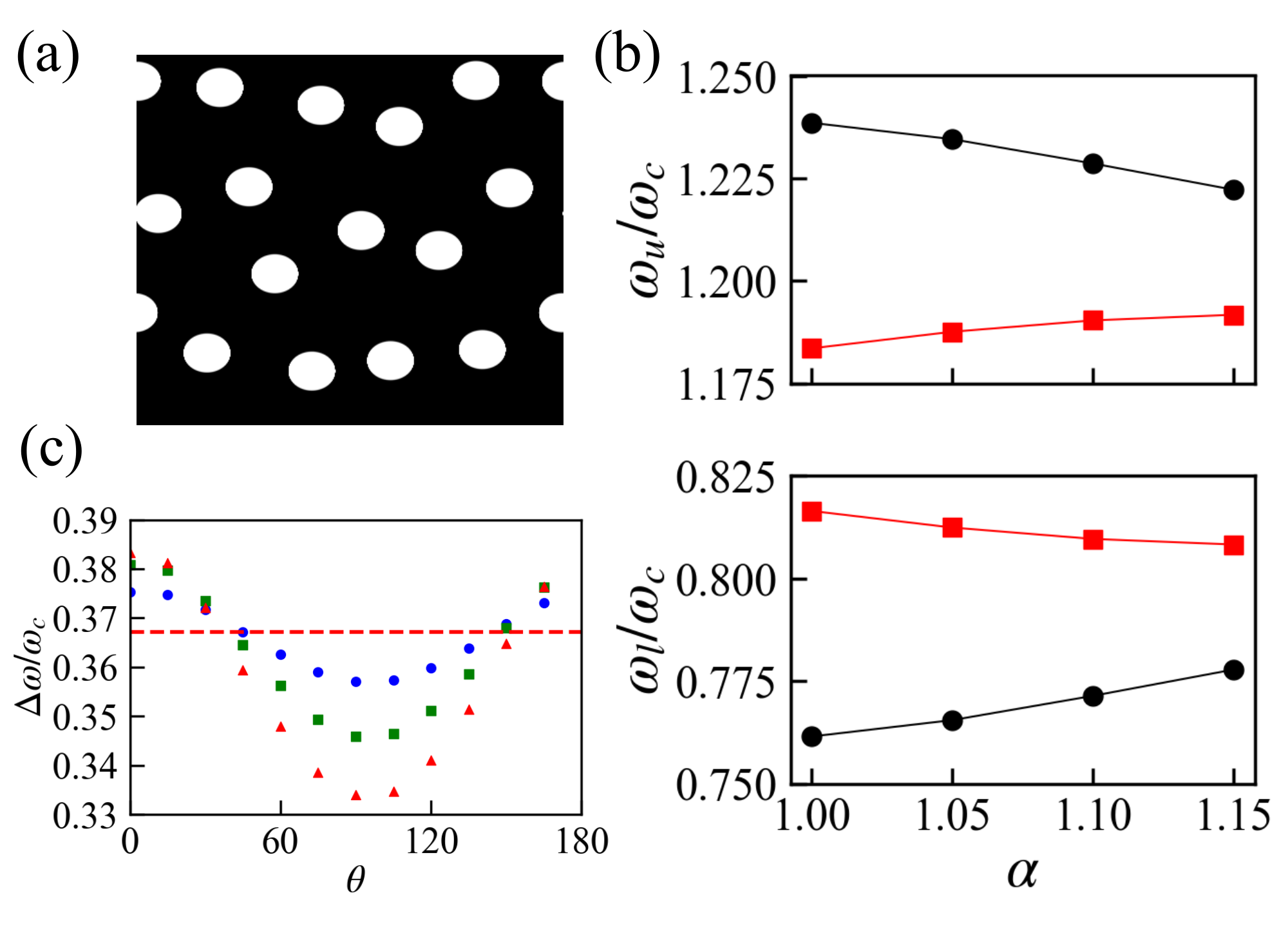}
  \caption{
    (a) Dielectric profile for the photonic crystal composed
    of the elliptrical rods with $R=0.2$,$\alpha=1.1$ and $\theta=0^{\circ}$ on the stealthy hyperuniform structure with $\chi = 0.41$.
    (b) Upper and lower band edges as a function of $\alpha$ for the elliptical rods with $\theta=0^{\circ}$.
    Red squares (black circles) represent the results for the stealthy hyperuniform point pattern (triangular lattice).
    (c) Relative PBG as a function of $\theta$ when $\alpha = 1.0$ (red dashed line), $1.05$ (blue circles), $1.1$ (green squares), and $1.15$ (red triangles).
  }
  \label{fig3}
\end{figure}
\begin{figure*}[hbt]
  \centering
\includegraphics[width=\linewidth]{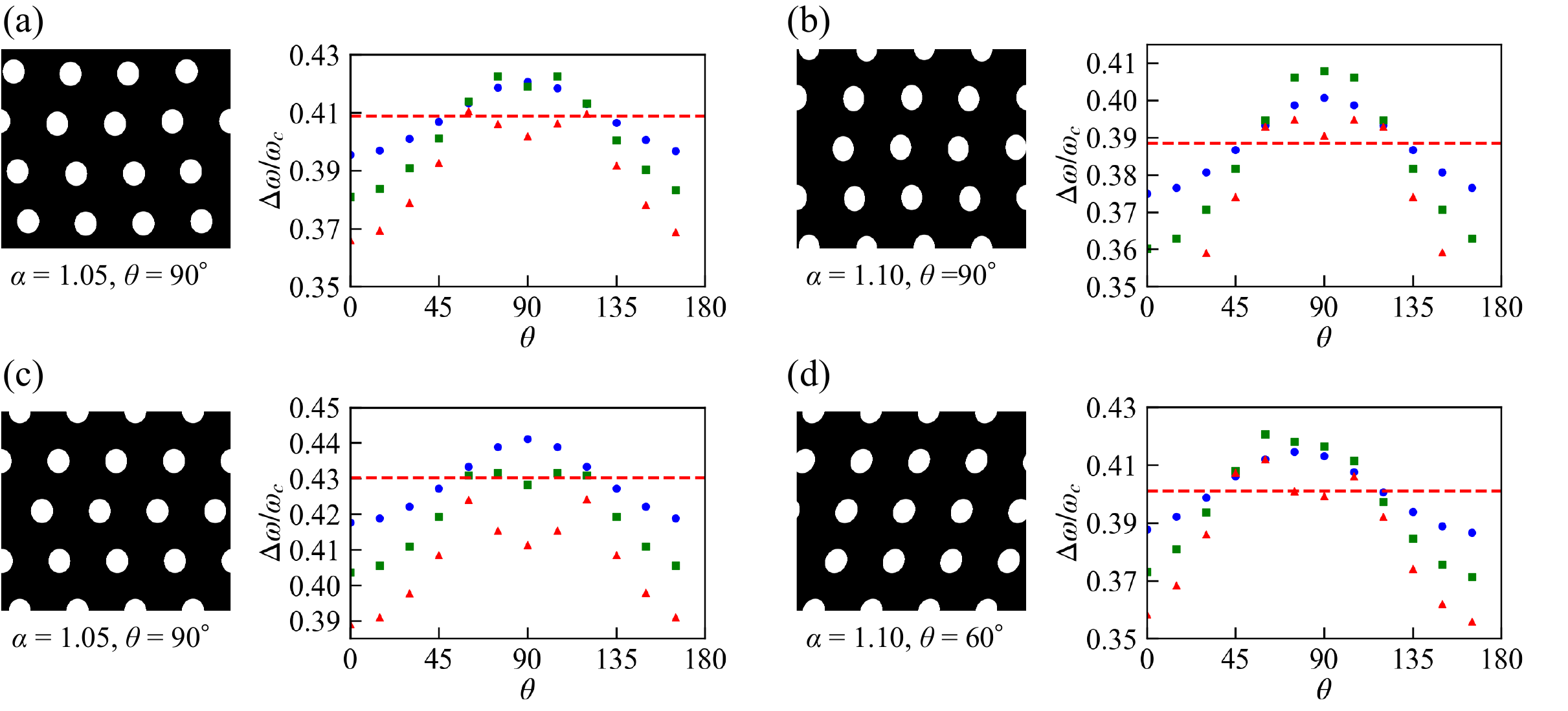}
  \caption{
  Dielectric profile (left) and the corresponding relative PBGs (right) for different configurations (a), (b), (c), and (d) with the same stealthiness parameter $\chi=0.63$. The relative PBGs are shown as a function of the rotation angle $\theta$ of the elliptical rods. The red dashed line denotes the relative PBG for cylindrical rods ($\alpha=1$), and blue circles, green squares, red triangles represent elliptical rods with $\alpha=1.05,1.10,1.15$, respectively.}
  \label{laxis2}
\end{figure*}

In this section, we consider photonic crystals composed of elliptical rods with $R=0.2$,
which are arranged according to a stealthy hyperuniform point pattern.
Such patterns are generated
by minimizing the objective function
starting from distinct random initial distributions.
We first examine simple systems
where all elliptical rods share the same deformation parameter 
$\alpha_i=\alpha$ and orientation angle $\theta_i=\theta$.
Distinct optical characteristics emerge in this system, in contrast to those of the periodic structure discussed above.
Figure~\ref{fig3}(a) shows the stealthy hyperuniform rod structure
with $\chi = 0.41$, $\alpha=1.10$ and $\theta=0^\circ$.
Using the standard plane-wave expansion method,
we obtain the photonic band structure.
The upper and lower edges of PBG are shown in Fig.~\ref{fig3}(b).
We find that, as $\alpha$ increases, $\omega_u$ shifts upward
while $\omega_l$ shifts downward.
Consequently, the relative PBG of the photonic crystal increases, 
indicating that the elliptical-rod geometry can enhance the PBG.
This behavior contrasts with that of the triangular-lattice system, where the PBG tends to shrink.
However, such enhancement is not always observed in the elliptical-rod systems.
Figure~\ref{fig3}(c) shows the relative PBG as a function of $\theta$.
We find that only for rotation angles $0\le\theta\lesssim 30^\circ$ and $150^\circ \lesssim \theta \le 180^\circ$, 
the PBG exceeds that of the cylindrical system with $\alpha=1$.
Our results demonstrate that the structural disorder inherent in
stealthy hyperuniform arrangements still 
permits a well-defined optimal orientation of the elliptical rods, which enhances the PBG.

The optimal relative photonic band gap should increase
with the stealthiness parameter $\chi$. 
Averaging over multiple realizations confirms the same trend reported in previous studies~\cite{Florescu09Jan}, 
indicating that the enhancement mechanism -- based on optimizing both rod orientation and aspect ratio -- 
remains robust against structural disorder. 
Therefore, introducing controlled ellipticity into stealthy hyperuniform structures provides 
a reliable approach in realizing photonic crystals with large and isotropic PBGs.

Here, we discuss the sample dependence of optical properties of
the photonic crystals composed of elliptical rods.
Here, we fix the stealthiness parameter at $\chi=0.63$.
For the elliptical rods with $\alpha=1.05, 1.1$, and $1.15$,
the relative PBGs of the photonic crystals
with four distinct stealthy hyperuniform point patterns are shown in Fig.~\ref{laxis2}.
We find that, in all cases,
the relative PBG exceeds that of the cylindrical case $(\alpha=1)$
within certain ranges of $\alpha$ and $\theta$, depending on the samples.
For example, in Fig.~\ref{laxis2}(a),
the relative PBG exceeds that of the cylindrical case
by approximately 1\% at rotation angles of $75^\circ \lesssim \theta \lesssim 105^\circ$,
when $\alpha = 1.05$ and $1.1$.
When $\alpha=1.15$, however,
the relative PBG is always smaller than that of the cylindrical rod ($\alpha=1$).
This behavior contrasts with the results in Fig.~\ref{laxis2}(b),
where the relative PBG exceeds that the cylindrical one for $\alpha\le 1.15$
and $60^\circ \lesssim \theta \lesssim 120^\circ$,
and increases with increasing $\alpha$.
Therefore, the sample dependence of the optical properties is pronounced.
In fact, the photonic crystals
shown in Figs.~\ref{laxis2}(a)-(d), exhibit distinct optical characteristics.

Finally, we aim to design a photonic crystal
composed of elliptical dielectric rods
such that its relative PBG is maximized.
To this end, we optimize the deformation parameter and orientation of
each elliptical rod, denoted as $\{\alpha_i, \theta_i\}$,
while keeping the rod positions fixed.
Here, we consider a stealthy hyperuniform pattern
shown in Fig.~\ref{fig3}(a) as an example.
\begin{figure}[hbt]
  \centering
  \includegraphics[width=\linewidth]{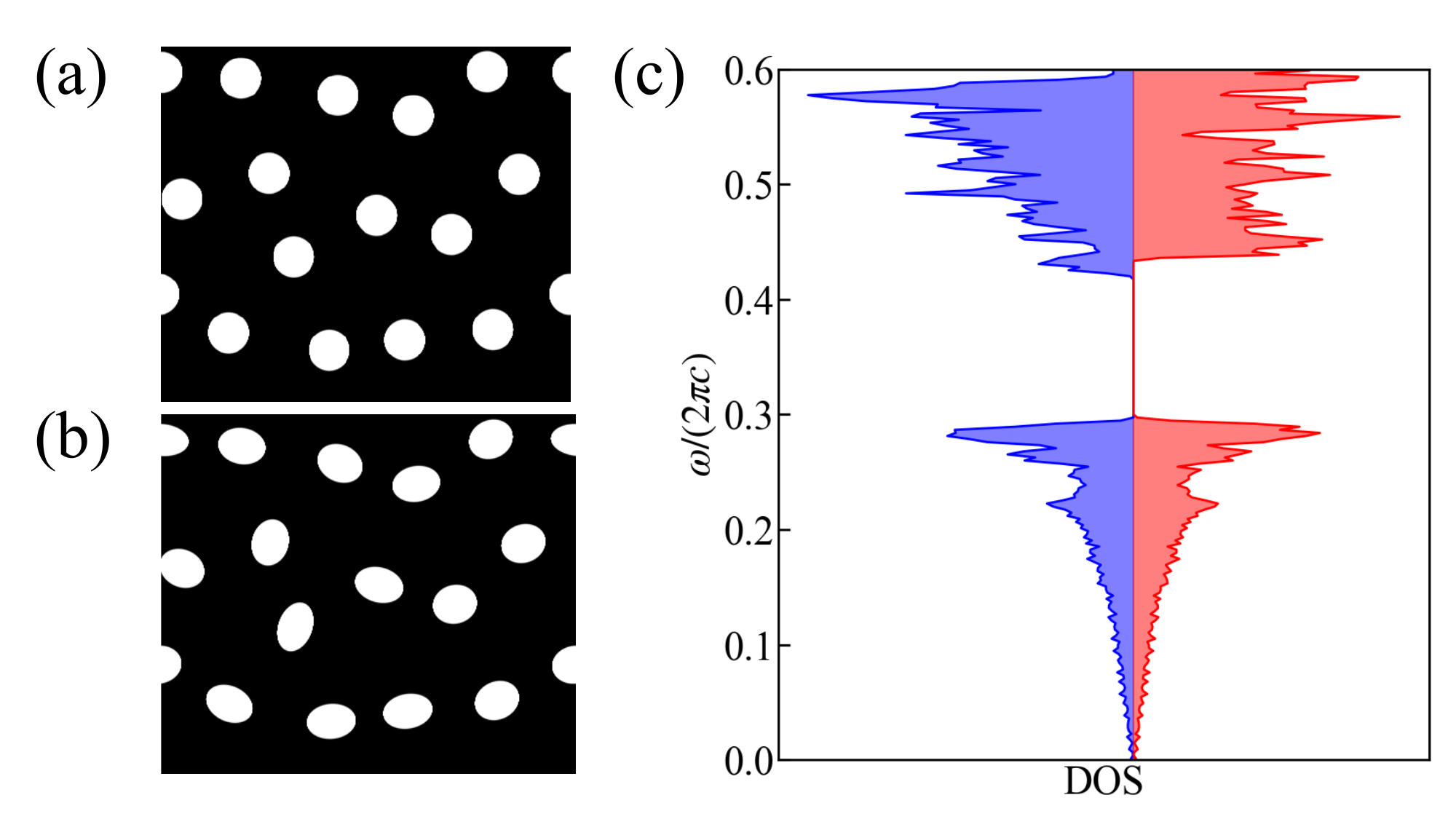}
  \caption{Dielectric profiles for the (a) cylindrical rod structure and (b) elliptical rod structure optimized with respect to $\{\alpha_i, \theta_i\}$.
    (c) DOS for the photonic crystals  with cylindrical (left) and elliptical rods (right). 
}
  \label{opt}
\end{figure}
Figures~\ref{opt}(a) and \ref{opt}(b) show the photonic crystal structures 
with cylindrical and optimized elliptical rods  
on the same stealthy hyperuniform point pattern.
The corresponding densities of states (DOS) are shown in Fig.~\ref{opt}(c).
We find that, in both cases, the PBG is located around  $\omega\sim 0.37\times (2\pi c)$.
These two DOS exhibit similar behavior in low-frequency region
while a slight difference appears above the PBG. 
Optimization of $\{\alpha_i, \theta_i\}$ enlarges the relative PBG from 36.7\% to 39.5\%,
indicating the effectiveness of geometric optimization.  
In the optimized structure, the deformation parameters $\alpha_i$ are distributed within a narrow range of $1.075 - 1.3$, 
while the orientation angles $\theta_i$ vary widely.

Only small-size calculations with $N=16$ are performed in this study.
We have found that the PBG of the photonic crystal arranged according to the stealthy hyperuniform structure does not exceed 
that of the triangular lattice.
Nevertheless, our findings indicate that
introducing controlled ellipticity and orientational optimization 
provides an effective strategy for enhancing and tuning the PBG 
in stealthy hyperuniform photonic crystals.

\newpage
\section{Conclusions}\label{sec_conclusion}
We have investigated two-dimensional photonic crystals
in which elliptical dielectric rods are arranged
according to stealthy hyperuniform point patterns.
Using the plane-wave expansion method to solve Maxwell’s equations,
we have analyzed their optical properties.
It has been found that photonic crystals composed of elliptical dielectric rods
can exhibit larger PBGs than those with cylindrical rods
when appropriate rotation angles and aspect ratios are chosen.
Furthermore, by examining the effect of the stealthiness parameter $\chi$,
we have elucidated that elliptical rod structures maintain their advantage over
cylindrical ones even in highly disordered arrangements.
This behavior contrasts with that observed in periodic lattices,
such as triangular and square arrays, where the introduction of ellipticity does not lead
to improvements in PBG.

It is worth exploring whether there exist dielectric rod geometries that
offer larger PBGs than the cylindrical and elliptical shapes.
As the structures studied in this work can be experimentally
tested in a variety of materials~\cite{Muller17,Amoah15,Aubry20,Tajiri20,Quan11,Kang18},
it is interesting to explore PBGs in samples other than silicon.
Such structures are expected to be applicable to devices
such as photonic waveguides that require large and
spatially isotropic PBGs~\cite{Lin00,Tokushima00,Milošević19}.

\begin{acknowledgements}
  We would like to thank N. Takemori for valuable discussions.
  Parts of the numerical calculations were performed
  in the supercomputing systems in ISSP, the University of Tokyo.
  This work was supported by Grant-in-Aid for Scientific Research from
  JSPS, KAKENHI Grants No.\ JP25K17327 (K.Y.),
  JP22K03525, JP25H01521, and JP25H01398 (A.K.).
  This work was partly funded by Hirose Foundation,
  the Precise Measurement Technology Promotion Foundation, and
  the Fujikura Foundation.
\end{acknowledgements}

\nocite{apsrev42Control}
\bibliographystyle{apsrev4-2}
\bibliography{example.bib}
\end{document}